\documentclass{pasj00}

\begin{document}
\SetRunningHead{S. Kato}{Mass and Spin of GRS 1915+105}
\Received{2004/07/04}
\Accepted{2004/07/12}

\title{
Mass and Spin of GRS 1915+105 Based on a Resonance Model of QPOs
}

\author{Shoji \textsc{Kato}%
  }
\affil{Department of Informatics, Nara Sangyo University, Ikoma-gun,
  Nara 636-8503}
\email{kato@io.nara-su.ac.jp; kato@kusastro.kyoto-u.ac.jp}


%

\KeyWords{accretion, accretion disks --- kiloherz quasi-periodic 
    oscillations --- relativity --- stars: individual (GRS 1915+105)
    --- warps} 

\maketitle

\begin{abstract}

We demonstrate that the four high-frequency QPOs observed in 
GRS 1915+105 can be interpreted as the oscillation modes on disks
that non-linearly and resonantly interact with a warp.
The warp is assumed to be a low-frequency global pattern on the disk.
This identification suggests that the relevant mass, $M$, and 
spin parameter, $a$, 
are $M=13$--$14 M_\odot$ and $a=0$--$0.15$. 

\end{abstract}

\section{Introduction}

GRS 1915+105 is the most extraordinary variable X-ray binary known.
It exhibits a rich set of time variability in X-rays.
So far, four high-frequency quasi-periodic  oscillations (HFQPOs) have been
observed in GRS 1915+105.
Their typical frequencies are 41 Hz, 67 Hz, 113 Hz, and 168 Hz 
(a review by McClintock, Remillard 2003).
(These QPOs are hereafter called, in turn, the 40 Hz, 67 Hz, 113 Hz, and
168 Hz QPOs, although the frequencies may not represent real ones.)
The 113 Hz and 168 Hz QPOs are a pair in the sense  that their 
frequency ratio is close to 2 : 3.
GRS 1915+105 is one of three black hole binaries (the others are
XTE J1550$-$564 and GRO J1655$-$40) with a pair of high-frequency QPOs whose 
frequency ratio is close to 2 : 3.

Time variabilities are good tools to investigate the structure of rotating
gas in a strong gravitational field of black holes (e.g., Kato 2001).
There are a few models proposed to explain HFQPOs.
Abramowicz and Klu\'{z}niak (2001) and Klu\'{z}niak and Abramowicz (2001) 
suggest that the pair HFQPOs are due to 
a parametric resonance, and that their frequency ratio must be commensurable,
e.g., 2 : 3.
The 67 Hz QPO, which is known to be rather stable, is suggested by Nowak
et al. (1997) to be a trapped g-mode oscillation resulting from the
presence of the maximum of the radial epicyclic frequency in a strong
gravitational field (Okazaki et al. 1987; Perez et al. 1997).

Recently, Kato (2004a, 2004b) proposed a resonance model of HFQPOs.
He assumed that a warp, which is a low-frequency
and global pattern, is present on axisymmetric disks.
Some of the disk oscillations then nonlinearly and resonantly couple with the
warp (Kato 2004a, referred hereafter to Paper I), and 
are amplified or damped (Kato 2004b, referred hereafter to Paper II).
The purpose of this letter is to examine, from the viewpoint of frequency,
whether the observed four HFQPOs of GRS 1915+105 can be explained 
by this resonance model, and how much the resulting mass and spin of 
the central source actually are.

\begin{figure}
  \begin{center}
    \FigureFile(80mm,80mm){figure1.eps}
  \end{center}
  \caption{
Feedback processes of nonlinear resonant interactions acting on
oscillations.
The original oscillations are characterized by $\omega$, $m$, and $n$.
Since the warp corresponds to a wave mode of $\omega=0$, $m=1$, and $n=1$,
the nonlinear interaction between them brings about 
intermediate modes of oscillations of $\omega$, $m\pm 1$, and $n\pm 1$.
To these intermediate oscillations, the disk resonantly responds at
a certain radius.
Then, the intermediate oscillations feedback to the original oscillations
after the resonance.
This feedback process amplifies or dampens
the original oscillations, since resonances are involved in the interaction 
processes.(After Paper I) 
  }
  \label{fig:1}
\end{figure}

\section{Spin-Mass Relations Derived from QPO Frequencies}

We consider a disk that is deformed by the presence of a warp.
The warp is a global one-armed ($m=1$) pattern with a low frequency.
When the disk is not deformed and axisymmetric, there are two kinds of
oscillation modes, i.e., the p-modes and g-modes (e.g., Kato et
al. 1998).
Because of the deformation,  the disk oscillations non-linearly couple 
with the warp.
Some of the coupled oscillations (we call them intermediate oscillations) can
resonate with the disk at some particular radii, depending on the modes 
of the oscillations.
These intermediate oscillations can further interact non-linearly with
the warp to feedback to the original oscillations (see figure 1).
The feedback to the original oscillations passing through a resonance
is of interest, since by energy exchange with the disk through the
resonance the amplitudes of the original oscillations change with time.

Figure 1 demonstrates schematically the case where the warp has no
precession.
This case is considered first.
Next, the case where the warp has a slow precession with a low frequency,
$\omega_{\rm p}$, is discussed.

\bigskip\noindent
(a) The case where warp has no precession

The analyses in Paper I show that there are two kinds of resonances,
i.e., horizontal and vertical.
The analyses further show that if we consider long-wavelength
oscillations, the resonances occur near to the radii where one of the
following relations is realized:
\begin{equation}
   \kappa=0, \quad \sqrt{2}\Omega_\bot-\Omega_{\rm K}=\kappa, \quad
   \Omega_{\rm K}=2\kappa, \quad \sqrt{3}\Omega_\bot-\Omega_{\rm K}=\kappa,
\label{1}
\end{equation}
where $\Omega_{\rm K}(r)$ is the (relativistic) Kepler frequency and
$\Omega_\bot(r)$ and $\kappa(r)$ are the frequencies of the
vertical and radial epicyclic oscillations, respectively.
The radii where one of the relations of equation (\ref{1}) is satisfied 
are denoted hereafter, in turn, as $r_0$, $r_\sqrt{2}$, $r_4$, and
$r_\sqrt{3}$, which are functions of the spin parameter, $a$.
In the case of the Schwarzschild metric ($a=0$), the radii satisfying
the above relations are, in turn,
\begin{equation}
  r_0=3r_{\rm g}, \quad r_\sqrt{2}=3.62r_{\rm g}, \quad
  r_4=4.0r_{\rm g}, \quad r_\sqrt{3} =6.46r_{\rm g},
\label{2}
\end{equation}
where $r_{\rm g}$ is the Schwarzschild radius, defined by $r_{\rm g}=
2GM/c^2$, $M$ being the mass of the central black hole.

The most interesting resonances are those at $r_4$, since they are
amplified (Paper II).
The resonances at $r_4$ are realized  by horizontal resonances of
g-mode oscillations (including the p-mode oscillations without node in
the vertical direction), and the frequencies of the resonant oscillations 
are $m^*\Omega_{\rm K}\pm\kappa$ at $r=r_4$, depending on the modes 
of oscillations, where $m^*$ is zero or an integer (Paper I and II).
In the case of one-armed ($m=1$) oscillations, oscillations of
$2(m^*\Omega_{\rm K}\pm\kappa)$ are also expected (Paper II).
The most prominent oscillations among them are those of 
$2(\Omega_{\rm K}-\kappa)$ and $2\Omega_{\rm K}-\kappa$
(or $\Omega_{\rm K}+\kappa$), which are 
$2\kappa$ and $3\kappa$ in the present case of no precession (Paper II).
Hence, we think that the oscillations of $2\kappa$ and $3\kappa$ at
$r_4$ are the observed 2 : 3 pairs of HFQPOs.
That is, we take $3\kappa$ at $r_4$ to be 168 Hz.
This gives a relation between the spin parameter, $a$, and the mass of the 
central object, $M$, which is shown in figure 2 by the curve labelled by 
168 Hz.

The resonant oscillations at other radii ($\kappa=0$,
$\sqrt{2}\Omega_\bot-\Omega_{\rm K}=\kappa$, and
$\sqrt{3}\Omega_\bot-\Omega_{\rm K}=\kappa$) are all damped when the warp
is steady (Paper II).
Among these damping oscillations, however, weakly damping ones
will be amplified and observable when the warp is in a growing stage.
In this sense, the resonant oscillations at the above radii are also worth
investigating.
Observations show occasional appearances of an $\sim 40$ Hz QPO and the
so-called 67 Hz QPO.
Considering that these frequencies are relatively low compared with the pairs, 
we assume that they are resonant oscillations at $r_\sqrt{3}$.
Then, they are all vertical resonances of g-mode oscillations (Paper I), and
their frequencies are $m^*\Omega_{\rm K}\pm\kappa$ at $r_\sqrt{3}$
(Paper I), where $m^*$ is again zero, or a positive integer.
Among the oscillations of $m^*\Omega_{\rm K}\pm\kappa$ at $r_\sqrt{3}$,
$\Omega_{\rm K}-\kappa$ is too low for our present interest.
Hence, as the next lowest frequencies of resonant oscillations, we
adopt $\kappa$ and $2\Omega_{\rm K}-\kappa$.
That is, we regard the 40 Hz and 67 Hz QPOs as resonant oscillations 
with frequencies $\kappa$ and $2\Omega_{\rm K}-\kappa$ at
$r=r_\sqrt{3}$.
Recent observations show that the frequency of the so-called 67 Hz
QPOs is 69.2$\pm$0.15 Hz (Strohmayer 2001).
Hence, we adopt here $\kappa_{r=r_\sqrt{3}}=40$ Hz and 
$(2\Omega_{\rm K}-\kappa)_{r=r_\sqrt{3}}=69$ Hz.
They give two independent relations between $a$ and $M$, which are 
also shown in figure 2 with labels of 40 Hz and 69 Hz.
In order to demonstrate how much the curves shift on the diagram when
the frequencies are changed, the curves of 41 Hz and 68 Hz are also
shown in figure 2.

Figure 2 shows that three curves of 168 Hz, 69 Hz, and 40 Hz cross near to 
a point, although not at a point.
As the curves of 68 Hz and 41 Hz show, curves move to the left of  
the diagram if the labelled frequency is increased.
That is, if we adopt 39.9 Hz instead of 40 Hz, while keeping the other 
frequencies unchanged at 168 Hz and 69 Hz, the three curves cross at a point, 
giving $a\sim 0$ and $M\sim 12.7 M_\odot$.
If we adopt $69.3$ Hz instead of 69 Hz, while keeping the others 
unchanged at 168 Hz 
and 40 Hz, the three curves cross at an another point, giving 
$a\sim 0.02$ and $M\sim 12.9 M_\odot$.
As another example, the frequency of the 67 Hz QPO is increased to 
70 Hz, while keeping the others unchanged at 168 Hz and 40 Hz.
Then, the relations among 
the three curves become as shown in figure 3.
This shows that three curves cross at a point if the frequency 40 Hz is
slightly increased to 40.2 Hz, while keeping the other frequencies unchanged 
at 168 Hz and 70 Hz.
This case gives $a\sim 0.04$ and $M\sim 13.1 M_\odot$.
Figure 3 further shows that the three curves cross at a point if the
frequency of the 168 Hz QPO is decreased to 167 Hz, giving $a\sim 0.06$
and $M\sim 13.3M_\odot$
These considerations show that a unique set of $a$ and $M$ 
is possible for suitable sets of three frequencies near to 
(168 Hz, 70 Hz, and 40 Hz).
The mass obtained in this way is consistent with $M=14\pm 4 M_\odot$ 
observationally derived by Greiner et al. (2001).

\bigskip\noindent
(b) The case where warp has precession with frequency $\omega_{\rm p}$

So far, evaluations of the spin parameter, $a$, and the mass, $M$, 
are made under the assumption that the warp has no precession.
When the central black hole has spin, however, the warp 
generally has a precession.
If the warp is precessing with frequency $\omega_{\rm p}$, the resonant 
radii are modified from those given by equation (\ref{1}).
The analyses in Paper I suggest that in the lowest order of
approximations, the resonance conditions $\Omega_{\rm K}=2\kappa$
and $\sqrt{3}\Omega_\bot-\Omega_{\rm K}=\kappa$,
for example, are changed, respectively, to 
\begin{equation}
   \Omega_{\rm K}=2\kappa\pm\omega_{\rm p} \quad {\rm and} \quad
   \sqrt{3}\Omega_{\bot}-\Omega_{\rm K}=\kappa\pm\omega_{\rm p},
\label{3}
\end{equation}
where both $\pm$ are possible, but we hereafter take the upper sign
(+), since this case has a better agreement with observations.
The frequencies of the pairs at the radius of $\Omega_{\rm
K}=2\kappa+\omega_{\rm p}$ are then $2\Omega_{\rm K}-2\kappa\ (=
2\kappa + 2\omega_{\rm p})$ and $2\Omega_{\rm K}-\kappa\ (=  
3\kappa + 2\omega_{\rm p}).$\footnote
{The case where the upper frequency is $\Omega_{\rm K}+\kappa 
(=3\kappa+\omega_{\rm p})$ is also possible.}
They are denoted by $\omega_{\rm l}$ and $\omega_{\rm u}$, respectively.
The frequency difference is still $\kappa$, but the ratio of
the pair frequencies is deviated from 2 : 3.
The frequencies corresponding to the 40 Hz and the
67 Hz oscillations are now the frequencies of $\kappa$ and 
$2\Omega-\kappa$ at a
radius of $\sqrt{3}_\bot-\Omega_{\rm K}=\kappa+\omega_{\rm p}$.

The case of $\omega_{\rm p}=0.5$ Hz with 168 Hz, 69 Hz, and 39 Hz is shown
in figure 4.
Since $\omega_{\rm p}=1.5\omega_{\rm l}-\omega_{\rm u}$, we have
$\omega_{\rm l}=2\omega_{\rm u}/3+2\omega_{\rm p}/3$.
Hence, the adoption of $\omega_{\rm u}=168$ Hz and $\omega_{\rm p}=0.5$ Hz
means that we have adopted $\omega_{\rm l}=112.3$ Hz.
In figure 4,  the three curves cross near to a point.
To make the three curves cross closer, keeping 168 Hz and 69 Hz
unchanged, the frequency 39 Hz should be slightly decreased as 38.9 Hz.
Then, the crossing point gives $a\sim 0.11$ and $M\sim 13.8 M_\odot$.
As an example of a larger $\omega_{\rm p}$, the case of
$\omega_{\rm p}=2.0$ Hz is shown in figure 5 with 168 Hz (and 175 Hz), 69 Hz, 
and 37.5 Hz.
This figure shows that the curves of the 40 Hz QPO and of the 67 Hz QPO
cannot touch each other unless the frequency of the 40Hz QPO is taken 
to be as low as 37.5 Hz when 70 Hz is adopted for the 67 Hz QPO.
(If a frequency lower than 70 Hz is adopted for the 67 Hz QPO, a frequency 
smaller than 37.5 Hz is required for the 40 Hz QPO.)
The figure further shows that the frequency of the 168 Hz QPO must be
higher than 175 Hz for the three curves to cross at a point.
 
\begin{figure}
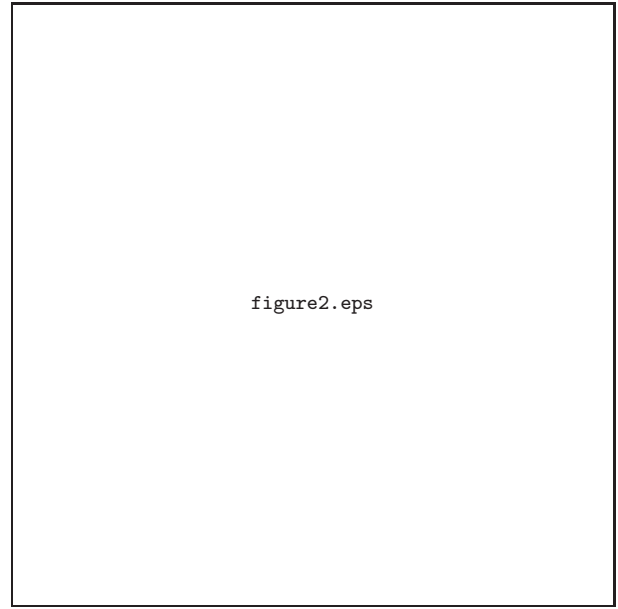

  \begin{center}
    \FigureFile(80mm,80mm){figure2.eps}
  \end{center}
  \caption{
Spin--mass relations derived from QPO frequencies, based on our
 resonance model.
The warp is assumed to have no precession.
Three curves crossing near to a point on the diagram are for 168 Hz, 
69 Hz, and 40 Hz.
In order to demonstrate how much the curves shift on the diagram when
 the frequencies are changed, the curves for 68 Hz and 41 Hz are also
shown.
If the frequency 40 Hz is slightly decreased to 39.9 Hz, the three
 curves cross at a point, giving $a\sim 0$ and $M\sim 12.7M_\odot$.
If  the frequency 69 Hz is slightly increased to 69.3 Hz, the three curves 
 cross at an another point, giving $a\sim 0.02$ and $M\sim 12.9M_\odot$.
}
  \label{fig:2}
\end{figure}

\begin{figure}
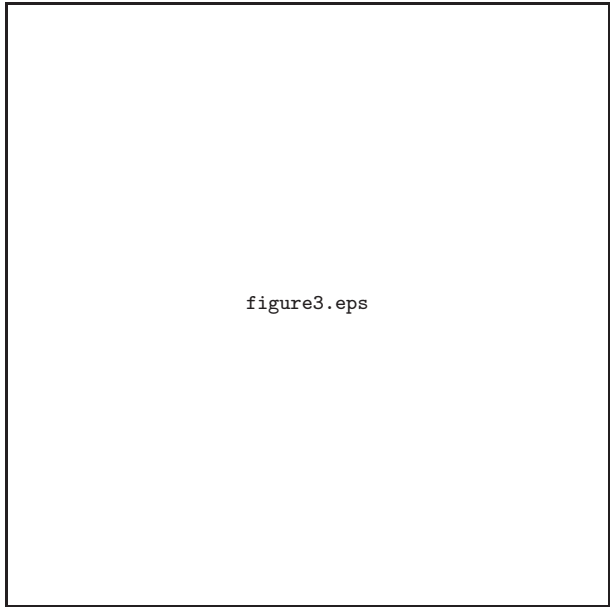

  \begin{center}
    \FigureFile(80mm,80mm){figure3.eps}
  \end{center}
  \caption{
Spin--mass relations derived from the QPO frequencies, based on our
 resonance model.
The warp is assumed to have no precession.
The three curves are for 168 Hz, 70 Hz, and 40 Hz.
These curves cross near to a point on the diagram.
 If the frequency, 40 Hz, is slightly increased to 40.2 Hz, the three curves
 cross at a point, giving $a\sim 0.04$ and $m\sim 13.1M_\odot$.
Similarly, if the frequency of the 168Hz QPO is decreased to 167 Hz,
 others being kept at 70 Hz and 40 Hz, the three curves cross at a point,
 giving $a\sim 0.06$ and $M\sim 13.3M_\odot$.
  }
  \label{fig:3}
\end{figure}

\section{Discussion}

The results given in section 2 suggest that the precession of the warp 
must be low.
That is, if $\omega_{\rm p}$ is as large as $\omega_{\rm p}=2.0$ Hz,
the requirement that the three curves cross at a point on the $a$--$M$
diagram imposes frequencies somewhat different from observations on the 
168 Hz, 67 Hz, and 40 Hz QPOs.
These considerations suggest that reasonable values of 
$\omega_{\rm p}$, $a$, and $M$ are $\omega_{\rm p}=0$--$0.7$ Hz,
$a=0$--$0.15$, and $M=13$--$14M_\odot$.

Let us discuss the possibility as to whether trapped c-mode oscillations 
are the warps required here.
Analyses by Silbergleit et al. (2001) show that the frequency 
of the trapped c-mode oscillations is $\sim 2$ Hz 
when $a\sim 0.1$ and $M\sim 14M_\odot$, 
which seems to be slightly larger than the 
precession frequency discussed in section 2.
Furthermore, the trapped region of the c-mode oscillations is rather
narrow in the radial extent (Silbergleit et al. 2001).
These results suggest that the trapped c-mode oscillations are not directly 
related to the warp.
GRS 1915+105 has many LFQPOs.
Some of them might be related to the warp.
The excitation of time-dependent low-frequency warps on disks is conceivable, if 
we remember that the global disk structure of GRS 1915+105 
occasionally changes by a phase transition between 
hard and soft states.
 
Nowak et al. (1997) suggested that the 67 Hz QPO is a trapped g-mode
oscillation.
In our present model, this is not the case, if we regard 113 Hz and 
168 Hz QPOs as being 2 : 3 pairs whose frequencies are close to 
$\Omega_{\rm K}$ and
$\Omega_{\rm K}+\kappa$ at the radius $r=r_4$.
Adoption of this model implies that $\kappa$ at $r=r_4$ is $\sim$56 Hz.
When $a=0$, this frequency is equal to $\kappa_{\rm max}$ (the maximum value 
of $\kappa$), since $r_4$ is the radius where 
$\kappa=\kappa_{\rm max}$.
This implies that when $a=0$, the 67 Hz QPO cannot be trapped g-mode
oscillations, since trapping requires that their frequencies
must be smaller than $\kappa_{\rm max}$.
When $a$ is large, the radius $r_4$ is outside of $r_{\rm max}$, and
$\kappa_{\rm max}$ is larger than $\kappa(r_4)$.
However, calculations show
that $\kappa_{\rm max}$ cannot become as
large as 67 Hz unless $a\sim 1$.
This means that the model of the trapped g-mode
oscillations and our present resonance model are incompatible.

In our  resonance model, the frequencies of the resonant oscillations
are not robust.
This is because the resonant oscillations are not radially trapped, 
and can propagate in the radial direction, while changing the
wavenumbers and frequencies.
This will be one reason why the QPOs have large $Q$-values.
Another reason for the large $Q$-values would be that different modes of
oscillations with different coupling paths have close frequencies.
If our resonance model represents real situations, some resonant
oscillations at $\sqrt{2}\Omega_\bot-\Omega_{\rm K}=\kappa$ may be
expected to be observed.

Finally, the azimuthal and vertical node numbers ($m$, $n$)
characterizing the resonant oscillations are summarized.
The pair oscillations at $r_4$ are p-mode or g-mode oscillations, coupling
horizontally with the warp.
The lower frequency oscillations of the pairs are the
oscillations of $m=1$ (one-armed oscillations).
They are the p-mode oscillation of $n=0$ or the g-mode oscillation with
arbitrary $n$, as discussed in Papers I and II.
The higher frequency oscillations of the pairs are (i) the g-mode
oscillations of $m=1$ or 2 with arbitrary $n$ or
(ii) the p-mode oscillation of $m=1$ or 2 with $n=0$.
The 40 Hz QPO and the 67 Hz QPO at $r=r_\sqrt{3}$ are due to the vertical 
resonance of the g-mode oscillations.
The 40 Hz QPO is the oscillation of ($m=0$, $n=2$) and the 67 Hz 
QPO is the oscillation of ($m=2$, $n=2$).

\bigskip
The author thanks Dr. Y. Kato for providing some references 
concerning observations.

\noindent

\begin{figure}
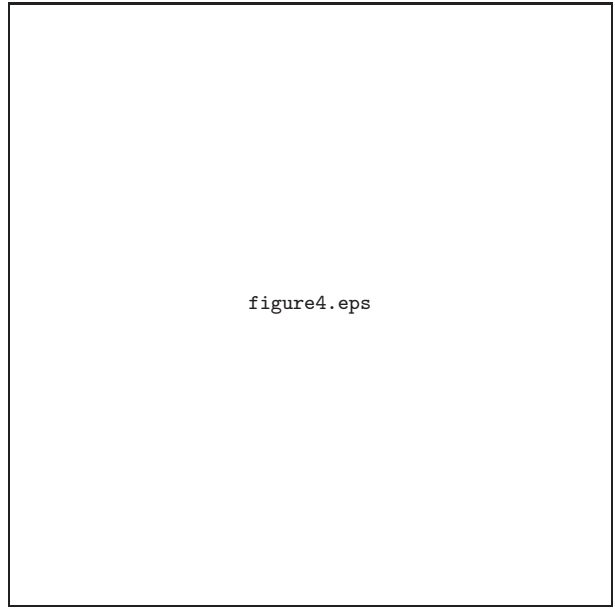

  \begin{center}
    \FigureFile(80mm,80mm){figure4.eps}
  \end{center}
  \caption{
Spin--mass relations derived from the QPO frequencies, based on our
 resonance model.
The warp is assumed to precess with a frequency of 0.5 Hz.
Three curves are for 168 Hz, 69 Hz, and 39 Hz.
If the frequency of 39 Hz is slightly decreased to 38.9 Hz, the three curves
 cross closer, giving $a\sim 0.11$ and $M\sim 13.8M_\odot$.
 }
  \label{fig:4}
\end{figure}

\begin{figure}
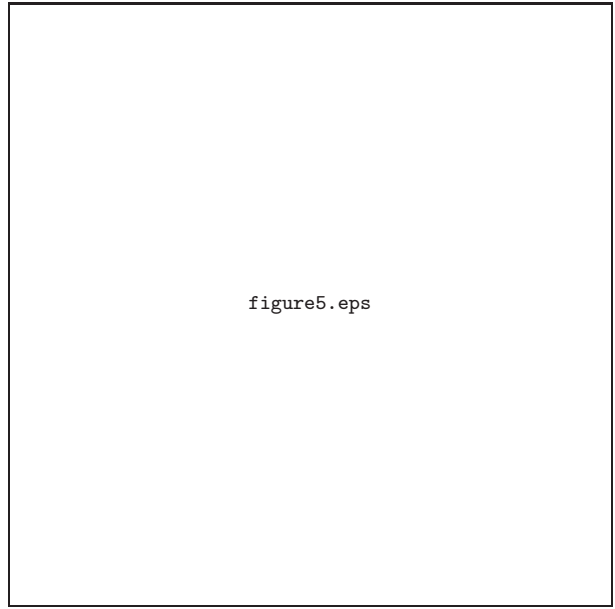

  \begin{center}
    \FigureFile(80mm,80mm){figure5.eps}
  \end{center}
  \caption{
Spin--mass relations derived from the QPO frequencies, based on our
 resonance model.
The warp is assumed to precess with a frequency of 2.0 Hz.
The four curves are for 175 Hz, 168 Hz, 69 Hz, and 37.5 Hz.
  }
  \label{fig:5}
\end{figure}



\end{document}